%Paper: hep-th/9406144
%From: dowker@a3.ph.man.ac.uk
%Date: Wed, 22 Jun 1994 12:22:50 +0100
%Date (revised): Tue, 28 Jun 1994 08:59:20 +0100

\input jytex.tex   % available from hep-th
%\draft
\typesize=10pt
\magnification=1200
\baselineskip=17truept
\hsize=6truein\vsize=8.5truein
%\leftmargin=1.25in
%\oddleftmargin=.5in
%\evenleftmargin=1.5in
\sectionnumstyle{blank}
\chapternumstyle{blank}
\chapternum=1
\sectionnum=1
\pagenum=0
% title style follows

\def\begintitle{\pagenumstyle{blank}\parindent=0pt\begin{narrow}[0.4in]}
\def\endtitle{\end{narrow}\newpage\pagenumstyle{arabic}}

% exercise style follows

\def\beginexercise{\vskip 20truept\parindent=0pt\begin{narrow}[10 truept]}
\def\endexercise{\vskip 10truept\end{narrow}}

% **************    my jyTeX abbreviations   *****************

\def\eql#1{\eqno\eqnlabel{#1}}
\def\ref{\reference}
\def\peq{\puteqn}
\def\pref{\putref}

\def\mgn{\marginnote}
\def\bex{\begin{exercise}}
\def\eex{\end{exercise}}
% *********************** My definitions ************************

\font\open=msbm10 %scaled\magstep1 % For VAX. Borde p195.
%\font\open=msym10 %scaled\magstep1 % For Arbortxt on PC
  % For Arbortxt on PC and VAX. Borde p199

\def\mbox#1{{\leavevmode\hbox{#1}}}
\def\hspace#1{{\phantom{\mbox#1}}}
\def\oR{\mbox{\open\char82}}
\def\oZ{\mbox{\open\char90}}

\def\rR{{\rm R}}

\def\al{\alpha}
 %in jyTeX
\def\brho{{\bmit\rho}} %in jyTeX
 %in jyTeX
\def\bka{{\bmit\kappa}}% in jyTeX
\def\be{\beta}

\def\de{\delta}

\def\ep{\epsilon}

\def\la{\lambda}

\def\om{\omega}
\def\Om{\Omega}

\def\si{\sigma}

\def\De{\Delta}
\def\vol{{\rm vol}}

\def\tr{{\rm tr\,}}

\def\det{{\rm det\,}}

     % Newline

\def\frac#1/#2{\leavevmode\kern.1em
\raise.5ex\hbox{\the\scriptfont0 #1}\kern-.1em/\kern-.15em
\lower.25ex\hbox{\the\scriptfont0 #2}}
\def\sfrac#1/#2{\leavevmode\kern.1em
\raise.5ex\hbox{\the\scriptscriptfont0 #1}\kern-.1em/\kern-.15em
\lower.25ex\hbox{\the\scriptscriptfont0 #2}}

\def\gtorder{\mathrel{\raise.3ex\hbox{$>$}\mkern-14mu
             \lower0.6ex\hbox{$\sim$}}}
\def\ltorder{\mathrel{\raise.3ex\hbox{$<$}|mkern-14mu
             \lower0.6ex\hbox{\sim$}}}

\def\semidirprod{\rlap{\ss C}\raise1pt\hbox{$\mkern.75mu\times$}}
\def\for{\lower6pt\hbox{$\Big|$}}
\def\fish{\kern-.25em{\phantom{abcde}\over \phantom{abcde}}\kern-.25em}

 %triple dot
 %double dot
 %double dot
%for small #1

\def\invo{^{\raise.18ex\hbox{${\scriptscriptstyle -}$}\kern-.06em 1}}
\def\inv#1{^{\raise.18ex\hbox{${\scriptscriptstyle -}$}\kern-.06em #1}}
%neg.power

\def\Dsl{\,\raise.18ex\hbox{/}\mkern-16.2mu D} %this one can be subscripted
\def\dsl{\raise.18ex\hbox{/}\kern-.68em\partial}
\def\slash#1{\raise.18ex\hbox{/}\kern-.68em #1}

\def\darr#1{\raise1.8ex\hbox{$\leftrightarrow$}\mkern-17mu #1} %backforarrowtop
\def\roughly#1{\ \lower1.5ex\hbox{$\sim$}\mkern-22.8mu #1\,}
\def\boxeqn#1{\vcenter{\vbox{\hrule\hbox{\vrule\kern3.6pt\vbox{\kern3.6pt
	\hbox{${\displaystyle #1}$}\kern3.6pt}\kern3.6pt\vrule}\hrule}}}
\def\boxit#1{\vbox{\hrule\hbox{\vrule\kern3pt
        \vbox{\kern3pt#1\kern3pt}\kern3pt\vrule}\hrule}}
 %small dalemb.
\def\dalemb#1#2{{\vbox{\hrule height .#2pt
        \hbox{\vrule width.#2pt height#1pt \kern#1pt
                \vrule width.#2pt}
        \hrule height.#2pt}}}

\def\ds{{|\!|}}         %double stroke
\def\cd#1{{}_{\ds #1}}  %lower covariant deriv.
 %upper covariant deriv.
\def\od#1{{}_{|#1}}     %lower ordinary  deriv.
\def\uod#1{{}^{|#1}}    %upper ordinary  deriv.

\def\noin{\noindent}

      %Connection
    %Connection'

\def\eg{{\it e.g. }}
\def\ie{{\it i.e. }}

\def\pa{\partial}

 %gives average <#1>
 %gives thermal average <<#1>>
   %gives bracket <#1|#2>
 %gives big bracket <#1|#2>
  %gives matrix element
%<#1|#2|#3>

  %

\def\3j#1#2#3#4#5#6{\left\lgroup\matrix{#1&#2&#3\cr#4&#5&#6\cr}
\right\rgroup}

\def\man{{\cal M}}

\def\m?{\mgn{?}}

%  *******************  Journal refs **********************

\def\aop#1#2#3{{\it Ann. Phys.} {\bf {#1}} (19{#2}) #3}

\def\cqg#1#2#3{{\it Class. Quant. Grav.} {\bf {#1}} (19{#2}) #3}

\def\jmp#1#2#3{{\it J. Math. Phys.} {\bf {#1}} (19{#2}) #3}
\def\jpa#1#2#3{{\it J. Phys.} {\bf A{#1}} (19{#2}) #3}

\def\np#1#2#3{{\it Nucl. Phys.} {\bf B{#1}} (19{#2}) #3}
\def\pl#1#2#3{{\it Phys. Lett.} {\bf {#1}} (19{#2}) #3}

\def\pr#1#2#3{{\it Phys. Rev.} {\bf {#1}} (19{#2}) #3}

\def\dmj#1#2#3{{\it Duke Math. J.} {\bf {#1}} (19{#2}) #3}

\def\jdg#1#2#3{{\it J. Diff. Geom.} {\bf {#1}} (19{#2}) #3}

\def\ma#1#2#3{{\it Math. Ann.} {\bf {#1}} ({#2}) #3}

% *******************   Main text *********************

\begin{title}
\vglue 20truept
\righttext {MUTP/94/14}
\righttext{hep-th/9406144}
\leftline{\today}
\vskip 100truept
\centertext {\Bigfonts \bf Effective actions with fixed points}
\vskip 15truept
\centertext{J.S.Dowker\footnote{Dowker@v2.ph.man.ac.uk}}
\vskip 7truept
\centertext{\it Department of Theoretical Physics,\\
The University of Manchester, Manchester, England.}
\vskip 60truept
\centertext {Abstract}
\vskip 7truept
\begin{narrow}
The specific form of the constant term in
the asymptotic expansion of the heat-kernel on an axially-symmetric space
with a codimension two fixed-point set of conical singularities is used to
determine the conformal change of the effective action in four dimensions.
Another derivation of the relevant coefficient is presented.
\end{narrow}
\vskip 5truept
\righttext {June 1994}
\vskip 100truept
\righttext{Typeset in \jyTeX}
\vfil
\end{title}
\pagenum=0
\section{\bf 1. Introduction and geometry}

In previous work [\pref{Dow4}] we have used the transformation of the
effective action,
obtained by integrating the conformal anomaly in two dimensions, to relate the
effective actions on regions of the two-sphere and plane. An extension to four
dimensions is technically feasible.
The relevant transformations when the manifolds have no boundary
were given some time ago [\pref{conft}], and, when a boundary exists, in
[\pref{DandS}.] (See also [\pref{DandW}]). However, as a preliminary, it is
necessary to extend the analysis to manifolds with conical, or vertex,
singularities and this is the object of the present paper.

The general method depends upon knowing the constant term in the
heat-kernel expansion, in this case the $C_2$ coefficient.
When the manifold possesses a singular O(2) fixed-point set of simple conical
type, the extra term in
$C_2$ has been determined by Fursaev [\pref{Fur2}]. In [\pref{Dow2}]
the expression was rederived and its conformal behaviour discussed. A further
analysis is presented in a later section.

We enlarge on the geometry. Let ${\cal N}$ be a totally geodesic submanifold of
$\man$, in particular a fixed-point set of codimension two. The global
symmetry group is O(2), generated by the circular Killing vector
$\pa/\pa\phi$, making $\man$ axially symmetric. Reperiodising
the polar angle, $\phi$, from $2\pi$ to $\be$ turns $\man$ into $\man_\be$, a
space with a simple conical singularity of `extent' ${\cal N}$. Close to
${\cal N}$, $\man_\be$ approximates to the product ${\cal C}_\be\times{\cal N}$
where ${\cal C}_\be$ is a cone of angle $\be$.

As hyper-cylindrical coordinates of a point $P$ on $\man$ we take (i) the
distance, $r$, from $P$ along the normal geodesic to its foot on ${\cal N}$,
(ii) the coordinates, $y^a$, of
this foot and (iii) the angle $\phi$, between the tangent to this geodesic, at
$r=0$, and some fiducial normal vector (parallely propagated along ${\cal N}$).
$\phi$ is the rotation angle `about'
${\cal N}$ and, with $r$, makes up the coordinates of the normal space.
Close to ${\cal N}$, $r$ and $\phi$ are the usual plane polar coordinates.

The metric of $\man$ is generally $g_{\mu\nu}dx^\mu dx^\nu$ and in cylindrical
coordinates is taken to be
$$ds^2= dr^2+f(r,y)d\phi^2+g_{ab}(y,r)dy^ady^b
\eql{metricm}$$where $f(r,y)$ is an even function of $r$ and tends to $r^2$
as $r\to0$. Also $g_{ab}(y,0)=h_{ab}(y)$, the metric on ${\cal N}$. The
region $r\le b$ forms a tube, $U{\cal N}$,\mgn{Seifert fibred?}
surrounding ${\cal N}$. If $g_{ab}(y,r)=h_{ab}(y)$, all the surfaces $r=$
const., $\phi=$ const. are totally geodesic.
\section{\bf 2. Heat-kernel coefficients}

The integrated heat-kernel expansion is written
$$K_\be(t)\sim{1\over(4\pi t)^{d/2}}\sum_{n=0,1/2,\ldots}^\infty C_nt^n =
{1\over(4\pi t)^{d/2}}\sum_{n=0,1/2,\ldots}^\infty (A_n+F_n)t^n
\eql{hker}$$ where $A_n$ is the usual volume integral, over $\man_\be$,
of a local, scalar density involving the curvature of $\man$. The $F_n$ are the
due to the conical singularity and are integrals over ${\cal N}$. In
particular, $F_2$ is given by
$$F_2=\int_{\cal N}f_2\,h^{1/2}d^{d-2}y =\int_{\cal N} f_2\,d\,\vol_{\cal N}(y)
\eql{eff2}$$
with the integrand,
$$
f_2={\pi\over B}(B^2-1)f_{2,1}+{\pi\over360 B}(B^4-1)f_{2,2}
\eql{coeff2}$$where $B=2\pi/\be$ and
$$\eqalign{
f_{2,1}=&\big({1\over6}-\xi\big)R+\la_1\big(\bka{\bf.}\bka
-2\tr(\bka{\bf.}\bka)\big)\cr
f_{2,2}=&\bigg(2R_{\mu\nu\rho\si}({\bf n}^{\mu}{\bf .n}^{\rho})({\bf n}^{\nu}
{\bf .n}^{\si})-R_{\mu\nu}{\bf n}^\nu{\bf .n}^\mu-{1\over2}\bka{\bf.}\bka\cr
&\hspace{*************************}
+\la_2\big(\bka{\bf.}\bka-2\tr(\bka{\bf.}\bka)\big)\bigg).\cr}
\eql{coeff22}$$
The constant coefficients, $\la_1$ and $\la_2$, of the conformally covariant
combination $\big(\bka{\bf.}\bka-2\tr(\bka{\bf.}\bka)\big)$ are unknown. The
${\bf n}^\mu_i$ are the normals to ${\cal N}$ and we may take
$n^\mu_i{n_\mu}_j=\de_{ij}$.

Although the extrinsic curvatures $\bka$ are zero for a fixed-point set, it is
necessary to retain them when making {\it general} conformal transformations.
\mgn{Is this true???} A derivation of (\peq{coeff22}) is given in section 4.

\section{\bf 3. Conformal transformations and the effective action}
Under a Weyl rescaling, $g_{\mu\nu}\to e^{-2\om}g_{\mu\nu}$,
$\man_\be\to{\overline\man_\be}$ and ${\cal N}\to\overline{\cal N}$. In order
to preserve the topology, the transformation function $\om(r,\phi,y)$ must have
period $\be$ in $\phi$.
In general, the O(2) symmetry is destroyed by the rescaling. $\overline{\cal
N}$ is a submanifold of ${\overline\man_\be}$, but not a totally geodesic one.
It has nonzero extrinsic curvatures.\mgn{Check this carefully!}

When evaluated on
${\cal N}$, or equivalently on ${\overline{\cal N}}$, $\om$ becomes
independent of the coordinates of the normal space so that, at the singularity,
${\overline\man_\be}\to{\overline{\cal C}_\be}\times{\overline{\cal N}}$ where
${\overline{\cal C}_\be}$ is a cone of angle $\be$ scaled by a factor
depending on its position in ${\overline{\cal N}}$. We could say that
${\overline\man_\be}$ has a `squashed' conical singularity.

We turn now to an evaluation of the change in the renormalised effective
action, $W_\rR$, under a conformal transformation for conformal coupling,
$\xi=1/6$.

The technique used in the present paper is that explained in
[\pref{Duff,DandS,Dow3,BandO}], involving the conformal transformation of
the constant term in the heat-kernel expansion. The general formula is
$$
W_\rR[e^{-2\om}g]-W_\rR[g]=\lim_{d\to
d'}(4\pi)^{-d/2}{C^{(d)}_{d'/2}[e^{-2\om}g]-C^{(d)}_{d'/2}[g]\over d-d'},
\eql{effct}$$
where $d'$ is the dimension of $\man_\be$ and $d$ is an arbitrary dimension. We
set $d'=4$.

This method, in contrast to that of integrating the conformal anomaly,
requires an application of finite conformal transformations
in $d$ dimensions. In the present instance there is little to choose between
the two methods so far as effort goes.

The total coefficient $C_2$ contains the standard volume term $A_2$, which
is dealt with in [\pref{conft}]. In the present paper we are
interested only in the effect of $F_2$.

The transformations needed are
$$\eqalign{
&R_{\mu\nu}{\bf n}^\nu{\bf .n}^\mu\to
e^{2\om}\big(R_{\mu\nu}{\bf n}^\nu{\bf .n}^\mu+(d-2)\om_{\mu\nu}{\bf n}^\mu
{\bf .n}^\nu+2\De_2\om+2(d-2)\De_1\om\big)\cr
\noalign{\vskip 4truept}
&R_{\mu\nu\rho\si}({\bf n}^{\mu}{\bf .n}^{\rho})({\bf n}^{\nu}{\bf .n}^{\si})
\to e^{2\om}\big(R_{\mu\nu\rho\si}({\bf n}^{\mu}{\bf .n}^{\rho})
({\bf n}^{\nu}{\bf .n}^{\si})+
2\om_{\mu\nu}{\bf n}^\mu{\bf .n}^\nu+2\De_1\om\big),\cr}
$$where $\om_{\mu\nu}=\om\cd{\mu\nu}+\om\od\mu\om\od\nu$,
and
$$\eqalign{
&\bka_{ab}\to e^{-\om}\big(\bka_{ab}+h_{ab}{\bf n}^\mu\om\od\mu\big),\cr
&\bka\to e^{\om}\big(\bka+(d-2){\bf n}^\mu\om\od\mu\big)\cr}
$$ for codimension 2. Also, for reference, $h^{1/2}\to e^{(2-d)\om}h^{1/2}$.

The change in $h^{1/2}f_{2,2}$, for example, is
$$\eqalign{
(4-d)\bigg(\om f_{2,2}+&2\om\uod{\mu}\om\od{\mu}+\om\cd{\mu\nu}{\bf n}^\mu
{\bf .n}^\nu+(1+d/2)\om\od\mu\om\od\nu{\bf n}^\mu{\bf .n}^\nu+\bka{\bf.}
{\bf n}^\mu\om\od\mu+\cr
&\la_2(d-2)\big(\om\od\mu\om\od\nu{\bf n}^\mu{\bf .n}^\nu+2\bka{\bf.}{\bf n}^
\mu\om\od\mu\big)\bigg)+O\big((4-d)^2\big)-2\widehat\De_2\om.\cr}
$$
all multiplied by $h^{1/2}$. $\widehat\De_2$ is the Laplacian intrinsic to
${\cal N}$.

It is then easy to show that the change in the integral (\peq{eff2}) is
proportional to $(d-4)$ so that, from (\peq{effct}), our final result can be
written
$$
W_\rR[e^{-2\om}g]-W_\rR[g]= \int_{\cal N}\De w\,d\vol_{\cal N}(y)
$$with
$$
\De w=-\om f_2+{2\pi\over B}\sum_{k=1}^2(B^k-1)\De w_k,
\eql{deltaw}$$
where, after setting $d=4$ and $\bka=0$,
$$
\De w_1=-{\la_1\over8\pi^2}\,\om\od\mu\om\od\nu{\bf n}^\mu{\bf .n}^\nu
\eql{ch1}$$
and
$$
\De w_2=-{1\over16\pi^2}\bigg(2\om\uod{\mu}\om\od{\mu}+
\om\cd{\mu\nu}{\bf n}^\mu{\bf .n}^\nu+
(2\la_2+3)\,\om\od\mu\om\od\nu{\bf n}^\mu{\bf .n}^\nu\bigg).
\eql{ch2}$$

The constants $\la_1$ and $\la_2$ remain undetermined. The derivation of the
corresponding terms in the
presence of a conventional boundary is somewhat complicated, involving either a
direct solution of the differential equations or the cleaner, but still
longish, functorial properties, [\pref{BandG}].

Note that, even if $\overline{\cal N}$ were an O(2) fixed-point subspace of
$\overline{\man}_\be$, there would still be a contribution from the
$-\bka{\bf.}\bka/2$ term in (\peq{coeff22}). In this case
$\om\od\mu{\bf n}^\mu$ is zero removing dependence on $\la_1$ and $\la_2$ in
the final answer.

\section {\bf 4. The heat-kernel expansion}
For completeness we present a derivation of (\peq{coeff2}) with
(\peq{coeff22}), similar to that in [\pref{Fur2}]. The basic idea is that,
close to ${\cal N}$,
the heat-kernel on $\man_\be$ is approximated by that obtained by a process of
{\it reperiodisation} from the heat-kernel on $\man$. This has been described
and used in earlier work [\pref{Dowcone, DowRSS, Dowcs}] where the Green
functions in some specific
curved spaces, \eg de Sitter and Schwarzschild, were considered.

We are interested in the integrated, diagonal $K_{\be}$
$$
K_{\be}(t)=\int_{\man_\be}K_{\be}(y,{\bf r};y,{\bf r},t)
$$
and its asymptotic expansion (\peq{hker}).

Because $\man_\be$ is identical to $\man$ off ${\cal N}$, the
local heat-kernel expansions of $K_\be$ and $K_{2\pi}$ will be the same in
$\man_\be-{\cal N}$. It is therefore sufficient for the
asymptotic expansion to write, thickening out ${\cal N}$ by setting
$\man_\be=U{\cal N}\cup\overline {U{\cal N}}$,
$$
K_{\be}(t)=\int_{U{\cal N}}K_{\be}(y,{\bf r};y,{\bf r},t)+
\int_{\overline {U{\cal N}}}K_{2\pi}(y,{\bf r};y,{\bf r},t)
\eql{split1}$$
valid up to terms exponentially small as $t\to0$.

Because of the O(2) symmetry, the Laplacian heat-kernel
$K_{2\pi}(y,r,\phi;y',r',\phi',t)$
depends on the polar angles through the difference $\phi-\phi'$ only, and the
approximation in the narrow tube $U{\cal N}$ is conveniently written as a
contour integral,
$$K_{\be,\de}(\phi-\phi',t)\approx\int_{A}K_{2\pi}(\al,t)\,
P(\al-\phi+\phi';\be,\de)\,d\al
\eql{perint}$$
where $P(\al;\be,\de)$ is the reperiodising function,
$$
P(\al;\be,\de)={e^{\pi i\al(2\de-1)/\be}
\over2i\be\sin\big(\pi\al/\be\big)}.
\eql{perfunc}$$ As is our wont, a phase
change $\de$, $0<\de\le1$, has been included. This will not be made
use of here but allows one to discuss fluxes running along the singularity.
The contour $A$ has two parts. In the upper half-plane it runs from
$(\pi-\ep)+i\infty$ to $(-\pi+\ep)+i\infty$ and in the lower
half-plane from $(-\pi+\ep)-i\infty$ to $(\pi-\ep)-i\infty$.

It is helpful to exhibit the arguments now,
\begin{ignore}
$$
K_{\be}(y,{\bf r};y',{\bf r}',t)\approx{1\over\be}\int_{A}K_{2\pi}(y,{\bf
r};y',
\brho',t){e^{2\pi i\al/\be}\over e^{2\pi i\al/\be}-e^{2\pi i(\phi-\phi')/\be}}
\,d\al\eql{Kbetacont}$$
\end{ignore}
$$K_{\be,\de}(y,{\bf r};y',{\bf r}',t)\approx{1\over2\be i}
\int_{A}\,K_{2\pi}(y,{\bf r};y',{\bf r}'_\al,t){e^{\pi i(\al-\phi+\phi')
(2\de-1)/\be}\over\sin\big(\pi(\al-\phi+\phi')/\be\big)}\,d\al.
\eql{perintd}$$
The `complex point' ${\bf r}'_\al$
has polar coordinates $(r',\al+\phi)$.

We also give the expression for $\de=1$, \ie no phase change,
$$K_{\be}(y,{\bf r};y',{\bf r}',t)\approx{1\over2\be
i}\int_{A}\,K_{2\pi}(y,{\bf r};
y',{\bf r}'_\al,t)\cot\big({\pi(\al-\phi+\phi')\over\be}\big)\,d\al
\eql{perintg}$$ where the symmetry of the contour under reversal of $\al$ has
been used and the fact that $K_{2\pi}(\phi,t)=K_{2\pi}(-\phi,t)$ by orientation
arguments.\mgn{is this true??} (The metric (\peq{metricm}) is unchanged under
reversal of $\phi$. If the cone were spinning, these arguments would have to be
revised.)

To isolate the effect of the singularities, the contour $A$ is deformed
to a small loop around the origin plus two infinite `vertical' curves,
labelled $A'$, which, to avoid problems, are taken to skirt the
origin. The small loop evaluates by the pole at $\al=0$ to $K_{2\pi}$
and so one has the split
$$K_{\be}\approx K_{2\pi}+K'_{\be}
\eql{split}$$
where $K'_{\be}$ is the effect of the singularity and is given by a formula
like (\peq{perintg}) but now over the $A'$ contour.

Effecting the split (\peq{split}), and combining with (\peq{split1}), one finds
$$K_\be(t)\sim{\be\over2\pi}K_{2\pi}(t)+
{1\over2\be i}\int_{A'}\int_{U{\cal N}}\,K_{2\pi}(y,{\bf r};y,{\bf r}_\al,t)
\,\cot\big({\pi\al\over\be}\big)\,d\al
\eql{tracecont}$$where
$K_{2\pi}(t)$ is the integrated kernel on the smooth manifold $\man$ and has
the standard asymptotic expansion. It will not be considered further.
The $\be/2\pi$ is a volume factor that reflects the O(2) symmetry. The
second term is the effect of the singularity. We denote it by $K'_\be(t)$.
The point $(y,{\bf r}_\al)$ is $(y,{\bf r})$ rotated through $\al$ about
${\cal N}$.

As explained in our earlier works, the contour $A'$ can be replaced
by a small clockwise loop around the origin, and we will imagine this to have
been done.

Donnelly [\pref{Donnelly}] has elucidated the asymptotic expansion of
$$
K_{2\pi}(\phi,t)=\int_{\man}\,K_{2\pi}(y,{\bf r};y,{\bf r}_\phi,t)=
{2\pi\over\be}\int_{\man_\be}\,K_{2\pi}(y,{\bf r};y,{\bf r}_\phi,t)
\eql{parhk}$$
and his results can be substituted
directly into (\peq{tracecont}) for, as we see, the complex activity takes
place in the normal space, the point $y$ of ${\cal N}$ being a
spectator.

It is clear from the classic results of Minakshisundaram and Pleijel that, up
to exponentially small terms, the integral in (\peq{parhk}) gets its value
from the fixed-point set ${\bf r}={\bf r}_\al$ \ie from ${\cal N}$, and so,
following Donnelly, for $\phi\ne0$,
$$
K_{2\pi}(\phi,t)\sim{1\over(4\pi t)^{(d-2)/2}}\sum_{n=0}^\infty t^n\int_
{\cal N}b_n(\phi,y)\,d\,\vol_{\cal N}(y).
\eql{Donn}$$
Substitution into (\peq{tracecont}), after setting $\phi\to\al$, gives
$$
K'_\be(t)\sim{1\over(4\pi t)^{(d-2)/2}}{1\over4\pi i}
\sum_{n=0}^\infty t^n\int_{\cal N}
\int_{A'}
b_n(\al,y)\cot\big({\pi\al\over\be}\big)\,d\al\,d\,\vol_{\cal N}(y)
\eql{contoursum}$$
and we now concentrate on the contour integral part of this equation
$$
b_n(y)={1\over4\pi i}\int_{A'}
b_n(\al,y)\cot\big({\pi\al\over\be}\big)\,d\al.
\eql{bcoeffn}$$

If $S(\phi)$ is the linear O(2) (actually SO(2)) action on the normal fibre,
the general form of the coefficients is, [\pref{Donnelly}],
$$ b_n(\phi,y)={1\over|\det(1-S)|}b_n'(\phi,y)
\eql{coeffk}$$where
$b_n'(\phi,y)$ is an O$(d-2)\times$O(2) invariant polynomial in the components
of $T\equiv(1-S)\invo$, the curvature of $\man$ and its covariant
derivatives.

We indicate the origin of (\peq{Donn}) and (\peq{coeffk}), [\pref
{Donnelly,CdeV}]. The local Minakshisundaram-\break Pleijel parametrix
expansion is
$$
K_{2\pi}(x;x',t)\sim{e^{-\Om(x,x')/2t}\over(4\pi t)^{d/2}}\sum_{n=0}^\infty
a_n(x,x')t^n
\eql{MandP}$$ where $\Om(x,x')$ is half the square of the geodesic distance
between $x$ and $x'$. This is substituted into (\peq{parhk}) and the integral
divided into one over ${\cal N}$, with coordinates $y$, and one over the normal
fibre, with coordinates ${\bf r}$. Because of the exponential cutoff the
integral is restricted to the tubular neighbourhood $U{\cal N}$. Transforming
the fibre coordinates to
normal coordinates, $x^i$, based at $y$, $2\Om$ becomes $\tilde
x\widetilde{(1-S)}(1-S)x$. (This is actually a diagonal form.) The remainder
of the integrand, including the volume
form and the $a_n$ coefficients, is expanded about the point $(y,0)$ and the
integrals over the $x^i$ extended to $\pm\infty$, again up to exponentially
small errors. Standard Gaussian integrals, familiar from perturbation theory,
then yield (\peq{coeffk}). The nontrivial O(2) tensor dependence
of $b'_n$ originates in the expansion of the volume factor. We remark that
Donnelly's general expression shows that there is always the contribution
$a_n(y,y)$ to $b'_n(\phi,y)$.

In particular, Donnelly calculates for any codimension (we include
the $\xi R$ coupling)
$$\eqalign{
b_0'(\phi,y)&=1\cr
b_1'(\phi,y)&=\big({1\over6}-\xi\big)R+{1\over6}R_i^i
+{1\over3}
R_{ijkl}T^{ji}T^{lk}+{1\over3}R_{ijkl}T^{jk}T^{li}-R_{ij}T^{ik}T^{jk}\cr}
\eql{Donncoeff}$$
where
$$
R_{ijkl}=R_{\mu\nu\rho\si}n^\mu_i n^\nu_j n^\rho_k n^\si_l\quad{\rm and}\quad
R_{ij}=R_{\mu\rho\nu\si}h^{\rho\si}n^\mu_i n^\nu_j.
\eql{normcurv}$$
$h_{\mu\nu}=g_{\mu\nu}-{\bf n}_\mu{\bf.n}_\nu$ is equivalent to the metric on
${\cal N}$, therefore
$$R_{ij}=R_{\mu\nu}n^\mu_i n^\nu_j-
R_{\mu\rho\nu\si}{\bf n}^\rho{\bf.n}^\si n^\mu_i n^\nu_j
$$
and
$$
R_i^i=R_{\mu\nu}{\bf n}^\mu{\bf. n}^\nu-
R_{\mu\rho\nu\si}({\bf n}^\rho{\bf.n}^\si)( {\bf n}^\mu{\bf. n}^\nu).
$$

For codimension two the calculation is simplified by noting that
$$R_{ijkl}={1\over2}R^{(2)}\big(\de_{ik}\de_{jl}-\de_{il}\de_{jk}\big),\quad
{\rm where}\quad
R^{(2)}=R_{\mu\nu\rho\si}({\bf n}^{\mu}{\bf .n}^{\rho})({\bf n}^{\nu}
{\bf .n}^{\si})
$$so
$$R_{ij}=R_{\mu\nu}n^\mu_i n^\nu_j-{1\over2}R^{(2)}\de_{ij}.
$$

Substituting into (\peq{Donncoeff}) gives
$$
b_1'(\phi,y)=\big({1\over6}-\xi\big)R+{1\over6}R_i^i
-R^{(2)}\tr T^2-{1\over6}R^{(2)}(\tr T)^2
+R_{\mu\nu}n^\mu_i n^\nu_j(T^2)^{ij}.
\eql{beeone}$$
With respect to normal coordinates,
$$S=\left(\matrix{\cos\phi&\sin\phi\cr-\sin\phi&\cos\phi\cr}\right);\qquad
T={1\over2(1-\cos\phi)}\left(
\matrix{1-\cos\phi&\sin\phi\cr-\sin\phi&1-\cos\phi\cr}\right)
$$and
$$T^2=-{1\over2(1-\cos\phi)}\left(
\matrix{\cos\phi&-\sin\phi\cr\sin\phi&\cos\phi\cr}\right).
$$
In terms of indices
$$T_{ij}={1\over2}\big(\de_{ij}+{\sin\phi\over1-\cos\phi}\ep_{ij}\big)
={1\over2}\big(\de_{ij}+\cot(\phi/2)\ep_{ij}\big).
\eql{indices}$$
The characteristic equation is $T^2-T=-{\bf 1}\,\det T=-\widetilde T T$.

One sees from the symmetry of the matrices that,
in (\peq{beeone}), $(T^2)^{ij}$ can be replaced by $\de^{ij}\,\tr T^2/2$.
Furthermore $\tr T=1$ giving,
$$
b_1'(\phi,y)={1\over6}\bigg((1-6\xi)R+\big(1+6\tr T^2\big)\big(
R_{\mu\nu}{\bf n}^\mu{\bf. n}^\nu-2R^{(2)}\big)\bigg).
\eql{beeone2}$$

These results give the explicit dependence on the angle $\phi$. If $\phi$ is
replaced by the complex angle $\al$, the expressions can
be substituted into the contour integral (\peq{bcoeffn}).
Noting that $|\det(1-S(\phi))|=2(1-\cos\phi)\to2(1-\cos\al)$ and
$\tr T^2=-1+1/(1-\cos\phi)$, from (\peq{bcoeffn}), (\peq{coeffk}) and
(\peq{beeone2}) we encounter the polynomials, [\pref{Dowcs1}],
$$P_k(\be,\de)={1\over\be i}\int_{A'}\,{1\over(1-\cos\al)^k}
{\cos(\pi\al(2\de-1)/\be)\over\sin\big(\pi\al/\be\big)}\,d\al.
\eql{perintf}$$

A routine residue calculation gives
$$\eqalign{
P_1(\be,\de)&= {1\over3}(B^2-1)-2B^2\si,\cr
\noalign{\vskip 5truept}
P_2(\be,\de)&={1\over90}(B^2-1)(B^2+11)
-{1\over3}B^2\si\big(B^2\si+2\big)\cr
\noalign{\vskip 5truept}
P_3(\be,\de)&= {1\over3780}(B^2-1)(2B^4+23B^2+191),\cr
\noalign{\vskip 5truept}
&\hspace{**}-{1\over90}B^2\si\big(B^4\si(2\si+1)-15B^2\si-24\big),\cr
P_4(\be,\de)&={1\over85050}(B^2-1)(B^2+11)(3B^4+10B^2+227),\cr
\noalign{\vskip 5truept}
&\hspace{**}-{1\over5670}B^2\si\big(B^6(3\si^2+4\si+2)+28B^4\si(2\si+1)
+294B^2\si+432\big),\cr}
\eql{eyes}$$where $\si=\de(1-\de)$. In this paper $\si=0$.

Combining terms in (\peq{beeone2}) we obtain (\peq{coeff2}) and
(\peq{coeff22}) for $\xi=0$, as promised.
\section{\bf 5. The general coefficient}
A typical term in the general coefficient $b_n'(\al,y)$ has the form
$$M_{ij\ldots kl}T^{ij}\ldots T^{kl}
\eql{gencoeff}$$
where $M_{\cdots}$ is an appropriate combination of the curvature,
$R_{\mu\nu\rho\si}$, its covariant derivatives and the normal vectors
$n^\mu_i$. Using (\peq{indices}), a parity argument, or the
symmetry of the $\al$-integral, shows that there can only be an even
number of $\ep$-symbols so (\peq{gencoeff}) reduces to a series of
contractions of $M$. From (\peq{indices}) it follows that each pair of
$\ep$-symbols will increase the order of the polynomials, $P_k$, by one.\break
{}From
dimensions, the maximum number of pairs in $b'_n$ equals $n$, an example being
$(R_{ijkl}T^{ij}T^{kl})^n$, and so the general form of (\peq{bcoeffn}) is
$$
b_{n-1}(y)={2\pi\over B}\sum_{k=1}^n P_k(B)G_{nk}
\eql{genform}$$
where the $G_{nk}$ are integrals over ${\cal N}$ of a local, scalar expression
constructed from the curvature of $\man$, its covariant derivatives and the
normals $n^\mu_i$. Form (\peq{genform}) was given by Fursaev
[\pref{Fur2}]. It allows one to set up a functorial method on the lines of
Branson and Gilkey [\pref{BandG}] for the determination of the coefficients.

\begin{ignore}
As an aside we note that in the case of the $b_1$ coefficient, it was not
necessary to use the symmetry of the $\al$-integral. The reduction of
(\peq{beeone}) to (\peq{beeone2}) was purely algebraic, something that will
not happen for the higher coefficients. For example, the possible combination
$********$ \mgn{find one} does not seem to reduce to traces algebraically.
\end{ignore}
\section{\bf 6. Comments}
Donnelly's expression can also be applied to the special case
where $\be$ is an integral part of $2\pi$, $\be=2\pi/q$, $q\in\oZ$. This would
involve a preimage summation which can be effected to yield the expected
answer, as mentioned by Fursaev [\pref{Fur2}].

A natural extension is to higher codimensions. However a simple process
similar to that of periodisation  does not appear to exist. If we  think of
codimension two as corresponding to a dihedral angle (with sides identified)
then codimension three corresponds to a trihedral corner, and the heat-kernel
for such a domain is unknown except for special cases.

Regarding the conformal transformation, in order to apply the result
(\peq{deltaw}) with confidence, it would be helpful to have an
independent check. This would entail finding two conformally related spaces,
with conformally related singular
fixed-point sets, on which one could determine the effective action, or at
least that part due to the singularities. A possibility is the Einstein
universe with a cosmic string [\pref{Dowkerccs}]. This is conformal to
$\oR^4$ with a string.

Another possibility is to check the heat-kernel coefficients (\peq{coeff22})
themselves. This can be done in any dimension since the coefficients are
universal. However it necessitates finding a tractable space with a deformed
conical submanifold \ie one with nonzero extrinsic curvatures.
\section {\bf Acknowledgements}
I would like to thank D.V.Fursaev for helpful remarks.

\newpage

% ************************** REFERENCES ************************
\vskip 5truept
\noin{\bf{References}}
\vskip 1truept
\begin{putreferences}
\vskip 1truept

\ref{Donnelly}{H.Donnelly \ma{224}{1976}161.}
\ref{Dow}{J.S.Dowker {\sl Effective actions on spherical domains},
{\it Comm.Math.Phys}, in the press.}
\ref{Dow4}{J.S.Dowker \cqg{11}{94}{557}.}
\ref{KCD}{G.Kennedy, R.Critchley and J.S.Dowker \aop{125}{80}{346}.}
\ref{Fur2}{D.V.Fursaev {\sl Spectral geometry and one-loop divergences on
manifolds with conical singularities}, JINR preprint DSF-13/94,
hep-th/9405143.}
\ref{HandE}{S.W.Hawking and G.F.R.Ellis {\sl The large scale structure of
space-time} Cambridge University Press, 1973.}
\ref{DandK}{J.S.Dowker and G.Kennedy \jpa{11}{78}{895}.}
\ref{ChandD}{Peter Chang and J.S.Dowker \np{395}{93}{407}.}
\ref{FandM}{D.V.Fursaev and G.Miele \pr{D49}{94}{987}.}
\ref{Dowkerccs}{J.S.Dowker \cqg{4}{87}{L157}.}
\ref{BandH}{J.Br\"uning and E.Heintze \dmj{51}{84}{959}.}
\ref{Cheeger}{J.Cheeger \jdg{18}{83}{575}.}
\ref{DandS}{J.S.Dowker and J.P.Schofield \jmp{31}{90}{808}.}
\ref{Dow2}{J.S.Dowker {\sl Heat kernels on curved cones},
hep-th/9606002, \cqg{}{94}{} to be published.}
\ref{Dow3}{J.S.Dowker \pr{39}{89}{1235}.}
\ref{BandO}{M.R.Brown and A.Ottewill \pr{D21}{85}{2514}.}
\ref{conft}{M.R.Brown \jmp{25}{84}{136}; R.J.Riegert
\pl{134}{84}{56}; E.S.Fradkin and T.Tseytlin \pl{134}{84}{187}; L.Bukhbinder,
S.Odintsov and A.Shapiro \pl{162}{85}{92}; L.Bukhbinder, V.P. Gusynin and
P.I.Fomin {\it Yad. Phys.} {\bf 44} (1986) 828 [Sov.J.Nucl.Phys.{\bf 44} (1986)
534]; J.S.Dowker \pr{D33}{86}{3150}.}
\ref{Dowcone}{J.S.Dowker \jpa{10}{77}{115}.}
\ref{DowRSS}{J.S.Dowker \pr{D18}{78}{1856}.}
\ref{Dowcs1}{J.S.Dowker \pr{D36}{87}{3095}.}
\ref{Dowcs}{J.S.Dowker {\sl Quantum field theory around conical defects} in
{\it The formation and evolution of Cosmic strings} edited by S.W.Hawking,
G.W.Gibbons and T.Vachaspati, Cambridge University Press, Cambridge 1990.}
\ref{BandG}{T.Branson and P.B.Gilkey {\it Comm. Partial Diff. Equns.} {\bf 15}
(1990) 245.}
\ref{Duff}{M.J.Duff, \np{125}{77}{334}.}
\ref{DandW}{A.Dettki and A.Wipf \np{377}{92}{252}.}
\ref{CdeV}{Y.Colin de Verdi\`ere {\it Compos. Mat.} {\bf 27} (1973) 159.}
%Reffs
\end{putreferences}
\bye